\documentclass[9pt,twocolumn,twoside]{osajnl}

\journal{pr} 

\setboolean{shortarticle}{false}

\title{Experimental verification of a coherence factorization law for quantum states}

\author[1,2,3]{Yi Zheng}
\author[4,5,6]{Cheng-Jie Zhang}
\author[1,2,3]{Zheng-Hao Liu}
\author[4]{Jian-Wei Shao}
\author[1,2,3,7]{Jin-Shi Xu}
\author[1,2,3,*]{Chuan-Feng Li}
\author[1,2,3]{Guang-Can Guo}

\affil[1]{CAS Key Laboratory of Quantum Information, University of Science and Technology of China, Hefei 230026, China}
\affil[2]{CAS Center For Excellence in Quantum Information and Quantum Physics, University of Science and Technology of China, Hefei 230026, China}
\affil[3]{Hefei National Laboratory, University of Science and Technology of China, Hefei 230088, China}
\affil[4]{School of Physical Science and Technology, Ningbo University, Ningbo 315211, China}
\affil[5]{State Key Laboratory of Precision Spectroscopy, School of Physics and Electronic Science, East China Normal University, Shanghai 200241, China}
\affil[6]{e-mail: zhangchengjie@nbu.edu.cn}
\affil[7]{e-mail: jsxu@ustc.edu.cn}
\affil[*]{e-mail: cfli@ustc.edu.cn}

\begin{abstract}
As a quantum resource, quantum coherence plays an important role in modern physics. Many coherence measures and their relations with entanglement have been proposed, and the dynamics of entanglement has been experimentally studied. However, the knowledge of general results for coherence dynamics in open systems is limited. Here we propose a coherence factorization law, which describes the evolution of coherence passing through any noisy channels characterized by genuinely incoherent operations. We use photons to implement the quantum operations and experimentally verify the law for qubits and qutrits. Our work is a step toward the understanding of the evolution of coherence when the system interacts with the environment, and will boost the study of more general laws of coherence.
\end{abstract}

\setboolean{displaycopyright}{true}

\begin{document}

\maketitle

\section{Introduction}

    Quantum coherence \cite{RMP} arising from the superposition principle is the key factor for quantum physics to deviate from classical one when describing particle systems, such as photons \cite{photons} which may have a classical wave theory, and purely matters in classical physics like atoms \cite{atoms} or ions \cite{ions}. The coherence leads to interference already known in classical optics, and also a purely quantum phenomenon in bipartite and multipartite systems, the quantum entanglement \cite{entangleRMP}. There have been several measures to quantify coherence, and they should fulfill the publicly accepted four conditions, non-negativity, monotonicity, strong monotonicity and convexity \cite{RMP}. One example of coherence measures is the $l_1$ norm of coherence $C_{l_1}$ introduced in Ref. \cite{Baumgratz2014}. Note that the coherence is dependent upon the reference basis we choose. A pure state has zero coherence if it is one of the basis states, but it may have coherence in other bases.

    Coherence and entanglement have many similar aspects \cite{Qi2017}. For instance, both of them can be treated as physical resources \cite{Baumgratz2014,RMP,resRMP}, and we often need to preserve the coherence against decoherence noises as well as entanglement in theories and experiments. The amount of entanglement has been quantified in many ways, such as the concurrence \cite{concur,entangleRMP}. Entanglement dynamics in open systems has been studied. Konrad \emph{et al}. provided a simple relation describing how the entanglement of two-qubit systems evolves as a whole system, under the action of an arbitrary noise channel on one of the components, and the dynamics of the entanglement becomes a very simple form fully captured by an entanglement factorization law \cite{Konrad}. Furthermore, using linear optical setups, Farías \emph{et al}. experimentally verified the entanglement factorization law under two quite different types of entanglement dynamics \cite{Farias}, and Xu \emph{et al}. characterized the bipartite entanglement under one-sided open system dynamics \cite{open1,open2,open3}, which includes pure and mixed initial two-photon state under the one-sided phase damping and amplitude decay channel \cite{Xu2009}. There have been more general results of the entanglement factorization law describing how bipartite high-dimensional entanglement evolves \cite{Gconcurrence}, and how multipartite entanglement \cite{multient} of a composite quantum system evolves when one of the subsystems undergoes a physical process \cite{Gour2010}.

    Compared with entanglement, the theoretical and experimental results on coherence dynamics in open systems are limited. We need some general law to determine its evolution equation to help us design the effective coherence preservation schemes. Theoretically, Hu \emph{et al}. introduced a framework of the evolution equation of coherence, and proved a simple factorization relation for the $l_1$ norm of coherence based on this framework, identifying the sets of quantum channels for which this factorization relation holds \cite{Hu2016}. The universality of this relation indicates that it applies to many other related coherence and quantum correlation measures \cite{coh1,coh2,coh3}. Experimentally, the coherence distillation involving strictly incoherent operations \cite{Xiong2021} and the achievement of the optimal state-conversion probabilities with stochastic incoherent operations \cite{Wu2020} have been demonstrated. To our knowledge, experimental results of general evolution law for coherence have never been reported before. Here, we provide a factorization law of a definition of coherence for quantum systems under a certain type of quantum operation, and experimentally verify it for qubits and qutrits using linear optics setup.

	\section{Theory}
	
	An arbitrary pure state in a $d$-dimensional Hilbert space $\mathcal{H}$ can be expressed as $|\psi\rangle=\sum_{i=1}^d a_i|i\rangle$. When all $a_i=1/\sqrt{d}$, the state is known as the maximally coherent state (MCS) \cite{Baumgratz2014} $|\psi^{+}\rangle=\sum_{i=1}^d|i\rangle/\sqrt{d}$, which should be one of the most coherent states in $\mathcal{H}$. The quantum operation is described by a set of Kraus operators $\{K_n\}$ satisfying $\sum_n K_n^\dagger K_n=I$. After the operation, the state $\rho$ becomes $\sum_n K_n\rho K_n^\dagger$. If each Kraus operator is diagonal in the reference basis $K_n=\operatorname{diag}(K_{n,1},K_{n,2},\ldots,K_{n,d})$, the operation is known as a genuinely incoherent operation (GIO), which has a property that an incoherent state is untouched after acting it, i.e., for any incoherent state $\delta$ a GIO $\Phi$ can preserve it $\Phi[\delta]=\delta$ \cite{RMP,GIO2016}. Moreover, all Kraus operators of GIO are diagonal in the reference basis \cite{GIO2016}. Therefore, the relations between GIO and incoherent operations (IO), strictly incoherent operations (SIO), maximally incoherent operations (MIO) are  GIO $\subset$ SIO $\subset$ IO $\subset$ MIO \cite{GIO2016}. In Refs.~\cite{Gour,Fan,Uhlmann,GME1}, the $G$-concurrence, which is the geometric mean of the Schmidt coefficients for pure states, has been proposed for full entangled states (i.e., with $d$ nonzero Schmidt coefficients) in $d\times d$ systems. The $G$-concurrence can describe entanglement factorization law in bipartite high-dimensional systems \cite{Gconcurrence}. Similarly, one can quantify the full coherence of a $d$-dimensional quantum state $\rho$ by defining the \emph{G}-coherence as 
	\begin{align}
		G(\rho)=d\prod_{i\ne j}|\rho_{ij}|^\frac{1}{d(d-1)},
	\end{align}
	which is $d$ times of the geometric mean of all $|\rho_{ij}|$ with $i\ne j$. When $d=2$ (qubit case), it becomes $2|\rho_{12}|$, the same as the $l_1$ norm of coherence $C_{l_1}(\rho)=\sum_{i\ne j}|\rho_{ij}|$ \cite{Baumgratz2014}. This definition means the full coherence is zero as long as there exists a zero off-diagonal element in the density matrix, and GIO will not keep the state unchanged if it has non-zero off-diagonal ones. Nevertheless, it will not bring coherence out of nothing. It is worth noticing that both the $G$-concurrence and the $G$-coherence are not designed for every entangled state and every coherent state, respectively. Consider a $d\times d$ Hilbert space, only entangled states with $d$ nonzero Schmidt coefficients contain positive $G$-concurrence, other entangled states with less than $d$ nonzero Schmidt coefficients do not contain any $G$-concurrence (such as any entangled two-qubit states in $3\times 3$ Hilbert space). The $G$-concurrence is designed for full entangled states in $d\times d$ systems. Similarly, for a $d$-dimensional system, the $G$-coherence is designed for full coherent states in $d$-dimensional system.
	
\begin{figure}[t!]
		\centering
		\includegraphics[width=7.8cm]{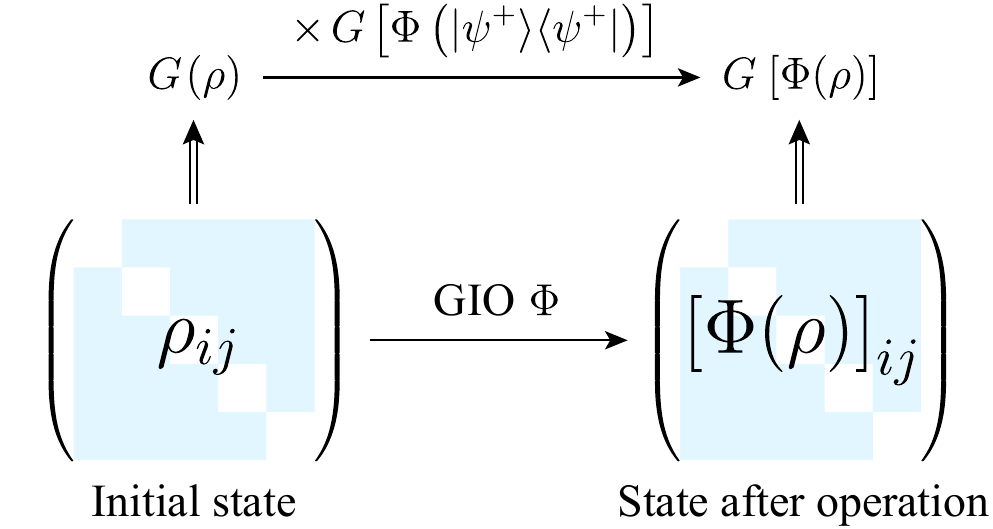}
		\caption{An illustration of the coherence factorization law under the genuine incoherent operation (GIO) $\Phi$. The coherence measure $G(\rho)$ is calculated from the off-diagonal elements of the density matrix. After GIO, the coherence is multiplied by $G\left[\Phi\left(|\psi^+\rangle\langle\psi^+|\right)\right]$, where $|\psi^+\rangle$ is the maximally coherent state (MCS).
		}
		\label{illustr}
	\end{figure}

	Here is the coherence factorization law: In a $d$-dimensional Hilbert space, if a GIO $\Phi$ acts on an arbitrary state $\rho$, pure or mixed, and produces the final state $\Phi\left(\rho\right)$, then the coherence measure $G$ of the final state is the product of the coherence of the initial state $\rho$ and the MCS after the operation $\Phi\left(|\psi^+\rangle\langle\psi^+|\right)$, that is,
	\begin{align}\label{theorem}
		G\left[\Phi\left(\rho\right)\right]=G\left(\rho\right)G\left[\Phi\left(|\psi^+\rangle\langle\psi^+|\right)\right].
	\end{align}
	We can see that the coherence of the final state is defined by two factors, one of which is solely determined by the initial state, and the other is only related to the operation. The operation term is reflected by the coherence of a state, which is similar to the Choi–Jamiołkowski isomorphism also known as the channel-state duality \cite{Datta2018}. An illustration of the principle is shown in Fig.~\ref{illustr}. The proof of the law uses a stronger theorem
	\begin{align}\label{stheorem}
	    \left[\Phi(\rho)\right]_{ij}=d\rho_{ij}\left[\Phi\left(\left|\psi^+\right\rangle\left\langle\psi^+\right|\right)\right]_{ij}=\rho_{ij}\left[\Phi(J_d)\right]_{ij}
	\end{align}
	(or in the Hadamard product form, $\Phi(\rho)=\rho\circ\Phi(J_d)$) for GIO about the factorization of each complex-valued matrix element, where $J_d=d|\psi^+\rangle\langle\psi^+|$ is the $d\times d$ all-ones matrix. The proof of this theorem uses the property of matrix product that the multiplication of the diagonal matrix $K_n$ on the left means the $i$th row is multiplied by $K_{n,i}$, and that of $K_n^\dagger$ on the right means the $j$th column is multiplied by $K_{n,j}^\ast$. So the elements of $\Phi(J_d)$ are
	\begin{align}
	    \left[\Phi\left(J_d\right)\right]_{ij}=\sum_n(K_n J_d K_n^\dagger)_{ij}=\sum_n K_{n,i}K_{n,j}^\ast.
	\end{align}
	Similarly,
	\begin{align}
	    \left[\Phi\left(\rho\right)\right]_{ij}=\sum_n(K_n\rho K_n^\dagger)_{ij}=\rho_{ij}\sum_n K_{n,i}K_{n,j}^\ast,
	\end{align}
	which proves Eq. \ref{stheorem}. Multiplying all the off-diagonal elements together and taking the modulus and the $d(d-1)$-th root yield Eq. \ref{theorem}.
	
    \section{Qubit experiment}
	
	The coherence of qubit $G(\rho)=2\left|\rho_{12}\right|$ is irrelevant to the diagonal elements of density matrix, so quantum state tomography \cite{Thew2002} at the $\{|D\rangle,|A\rangle\}$ and $\{|L\rangle,|R\rangle\}$ base is enough to yield this value
	\begin{align}\label{tomog}
	G=\sqrt{\langle\sigma_x\rangle^2+\langle\sigma_y\rangle^2},
	\end{align}
	where $\sigma_x$ and $\sigma_y$ are Pauli operators. In our experiments, we use the path degree of freedom (DOF) of photons as the system, and the polarization is the auxiliary DOF. The initial state is prepared as $\left|\psi\right\rangle=\sin2\theta_1\left|1\right\rangle+\cos2\theta_1\left|2\right\rangle$, and the Kraus operators of the operation we use are $K_1=\operatorname{diag}\left(\sin2\theta_2,\cos2\theta_2\right),K_2=\operatorname{diag}\left(\cos2\theta_2,\mathrm{i}\sin2\theta_2\right)$, which are suitable for implementing in an optical setup. When $\theta_1=$ 22.5°, $\left|\psi\right\rangle$ is the MCS.
	
	\begin{figure}[t!]
		\centering
		\includegraphics[width=8.8cm]{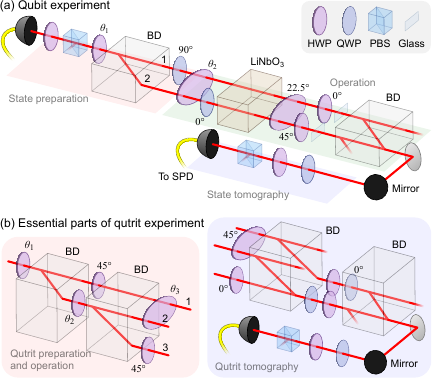}
		\caption{(a) The setup of the qubit coherence factorization law verification experiment. The photons are prepared as the initial state using a half-wave plate (HWP) with its fast axis at angle $\theta_1$ and a beam displacer (BD). An HWP at $\theta_2$ controls the operation, and two quarter-wave plates add a phase factor $i$ at Path 2. A thick lithium niobate ($\textrm{LiNbO}_\textrm{3}$) crystal destroys the coherence between the horizontally ($H$) and vertically ($V$) polarized component, and several HWPs erase the difference between the two polarizations. Two thin glass plates compensate for the phase difference between two paths before they are merged by another BD. A QWP, an HWP and a PBS perform state tomography and the photons are counted by a single-photon detector (SPD). (b) The essential parts of the qutrit experiment. The initial state is prepared using three HWPs and two BDs. The first HWP is at $\theta_1=$ 17.6° so that $\sin2\theta_1\approx1/\sqrt{3}$. The HWP at $\theta_3$ acts at the two paths on the top, and another HWP switches the photons at $V$ polarization at the bottom path to $H$ one. The tomography part is after the 22.5° HWP. Wave plates, two BDs and a PBS are used to project the photonic state to eigenstates of the Gell-Mann matrices.
		}
		\label{qubitsetup}
	\end{figure}
	
	\begin{figure}[t!]
		\centering
        \includegraphics[width=8.2cm]{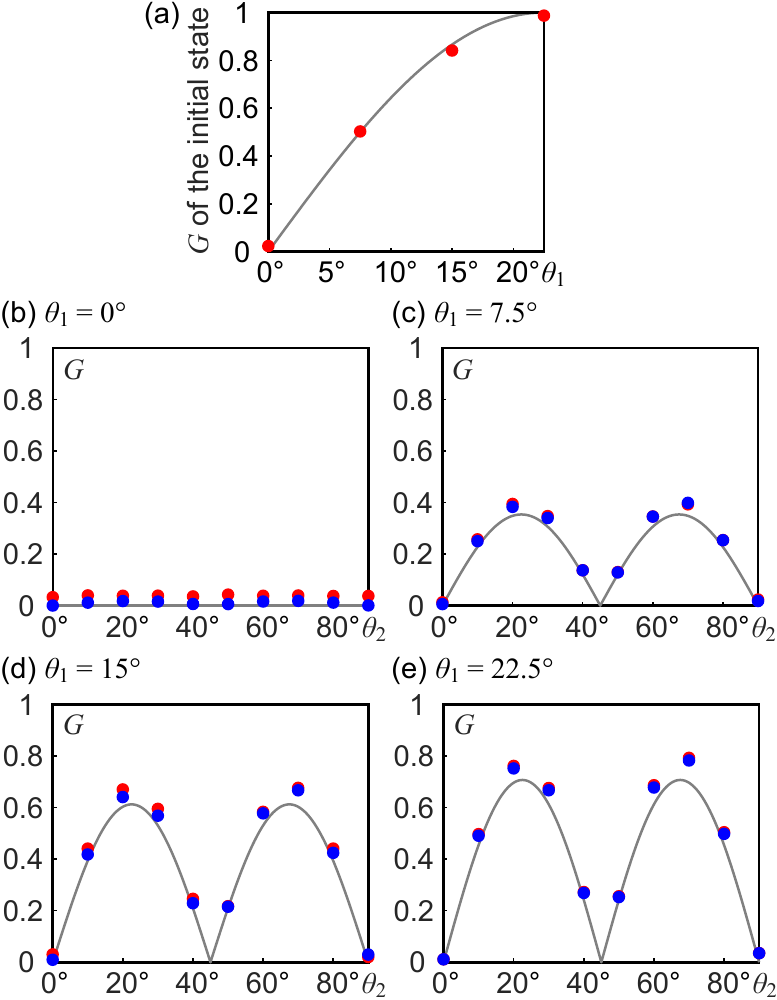} 
        \caption{Experimental results for qubit experiment. (a) The measured coherence $G$ for different initial states controlled by $\theta_1$ and the theoretical curve $\sin4\theta_1$. (b)–(e) The measured $G$ of the final state after the operation with different $\theta_1$ values (red dots), the product of the measured $G$ of the initial state and the measured $G$ of the maximally coherent state (MCS, when $\theta_2=$ 22.5°) after the operation (blue dots), and the theoretical curves $\left|\sin4\theta_1\sin4\theta_2\right|/\sqrt{2}$ (gray curves). The error bars from the $3\sigma$ deviations of the Poisson distribution are all smaller than the size of the dots and are thus omitted.}
        \label{qubitres}
    \end{figure}
	
	The experimental setup of qubit is shown in Fig.~\ref{qubitsetup} (a). An attenuated pulsed 808 nm light from a laser source is collimated into a single-mode fiber and prepared at the horizontal polarization $\left|H\right\rangle$ by a half-wave plate (HWP) and a polarizing beam splitter (PBS) after emitting from the collimator. Then an HWP with its fast axis at angle $\theta_1$ and a beam displacer (BD) produces the initial state $\left|\psi\right\rangle$, while the two paths have different polarizations. The HWP at angle $\theta_2$ acting on both paths changes the amplitudes at $H$ and $V$ polarization. A quarter-wave plate (QWP) at 0° is inserted at Path 2 to realize the phase factor $\mathrm{i}$ in $K_2$. To compensate for the optical path difference, another QWP is inserted at Path 1 right after the first BD. Then a thick lithium niobate ($\textrm{LiNbO}_\textrm{3}$) crystal introduces an optical path difference between the $H$ and $V$ component larger than the coherence length of the light source from the birefringence effect, and thus destroys the coherence between the two components \cite{thick}. In order to eliminate the polarization difference, an HWP at 22.5° converts the $H$ and $V$ polarization to diagonal ($D$) and anti-diagonal ($A$) one respectively. Two HWPs and the second BD select the $H$ component and merge the two paths together, converting the path information into polarization while discarding half the photons. Before merging, two thin glass plates are inserted at the two paths to compensate for the residual phase difference. Then the photons pass a tomography device consisting of a QWP, an HWP and a PBS, before being coupled into another single-mode fiber, sent to a single-photon detector (SPD), and counted by a computer. Dark counts are subtracted to increase the signal-to-noise ratio.
	
	We measured $G\left[\Phi\left(\left|\psi\right\rangle\left\langle\psi\right|\right)\right]$ with different $\theta_1$ and $\theta_2$ values using Eq.~\ref{tomog}. Then we remove the optical elements for the operation process from the 90° HWP to $\textrm{LiNbO}_\textrm{3}$ and directly performed tomography on the initial states with different $\theta_1$ values. The relation between $G$ of the initial state as well as the theoretical curve $\sin4\theta_1$ is plotted in Fig.~\ref{qubitres} (a). In Fig.~\ref{qubitres} (b)--(e), we plotted $G$ of the final states in red dots, the product at the right hand side of Eq. \ref{theorem} in blue dots, and the theoretical curves $\left|\sin4\theta_1\sin4\theta_2\right|/\sqrt{2}$ with different $\theta_1$ values. The product values are close to the directly measured $G$ values, verifying the factorization law.
    
    \section{Qutrit experiment}
	    
    \begin{figure}[t!]
		\centering
		\includegraphics[width=8.8cm]{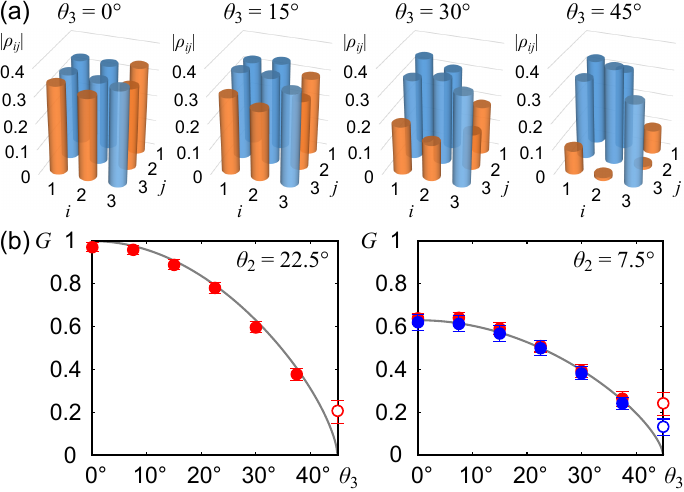}
		\caption{(a) The moduli of density matrix elements $|\rho_{ij}|$ from the state tomography of qutrit maximally coherent state (MCS) after the phase damping operation when $\theta_3=$ 0°, 15°, 30° and 45°. (b) The measured $G$ of the final qutrit state after operation (red dots) and the theoretical curves $G(\rho)\left|\sin4\theta_2\cos2\theta_3\right|^{2/3}$ when $\theta_2=$ 22.5° (MCS), where $G(\rho)=1$, and $\theta_2=$ 7.5°, where $G(\rho)=2^{-2/3}$, with the products of $G$ values of the initial state and the MCS after operation (blue dots). When $\theta_3=$ 45°, the value deviates from zero from the error amplification when taking cubic roots, and the dots are replaced by hollow circles. The error bars correspond to $3\sigma$ deviations of the Poisson distribution.
		}
		\label{qutritd}
	\end{figure}
	
    We use three path DOFs to study the qutrit scenario, where the initial state is $\left|\psi\right\rangle=\sqrt{1/3}\left|1\right\rangle+\sqrt{2/3}\cos2\theta_2\left|2\right\rangle+\sqrt{2/3}\sin2\theta_2\left|3\right\rangle$ and we design a phase damping operation between two of the paths and the other one. The corresponding Kraus operators are $K_1=\operatorname{diag}(\cos2\theta_3,\cos2\theta_3,1)$ and $K_2=\operatorname{diag}(\sin2\theta_3,\sin2\theta_3,0)$. Designing a true phase damping operation on all the paths is more complex as the polarization DOF as an auxiliary is two-dimensional. When $\theta_3=$ 0°, the coherence is preserved while it is completely destroyed when $\theta_3=$ 45°. The experimental setup is similar to the qubit experiment except for an additional path, a different wave plate setup to realize the operation, and a different tomography method. The essential parts are shown in Fig.~\ref{qubitsetup} (b). $\theta_2$ and $\theta_3$ are the angles of the wave plates controlling the initial state and the operation respectively. At the tomography process, wave plates, two BDs and a PBS are needed to project the quantum state to 15 eigenstates of 8 Gell-Mann matrices \cite{Gell-Mann,Thew2002,Liu2021}. The density matrix is calculated from the averages of these Hermitian operators. See Appendix for more details. Then we can obtain the coherence via $G(\rho)=3\sqrt[3]{\left|\rho_{12}\rho_{13}\rho_{23}\right|}$. However, if some of the off-diagonal elements should be zero while others are non-zero, a small value from the experimental errors will cause the measured $G$ to be significantly larger than zero. For example, in an experiment, if the moduli of the three measured off-diagonal elements are 0.01, 0.2 and 0.2, we have $G\approx0.22$. So, when the theoretical $G=0$, the experimental value will be inaccurate unless most of the off-diagonal elements are zero. 
    
    We choose $\theta_2=$ 22.5° (MCS) and 7.5° (another initial state) and take different $\theta_3$ values to perform state tomography. To show the impact of the phase damping operation on the moduli of density matrix elements, we present $|\rho_{ij}|$ values from our experiment in Fig.~\ref{qutritd} (a) when the initial state is MCS and $\theta_3=$ 0°, 15°, 30° and 45°. The off-diagonal element $\rho_{13}$ and $\rho_{23}$ decay as $\theta_3$ increases, while $\rho_{12}$ and the diagonal elements are roughly unchanged. The coherence values of the initial states are taken when $\theta_3=$ 0°. The measured $G$ values of final states, the products (not applicable for the MCS) and the theoretical curves $G(\rho)\left|\sin4\theta_2\cos2\theta_3\right|^{2/3}$ are plotted in Fig.~\ref{qutritd} (b). The coherence decays with the increase of $\theta_3$. When $\theta_3=$ 45°, the coherence should become zero but the experimental value is inaccurate as we have stated before, and the errors calculated from the $3\sigma$ deviation of the Poisson distribution at the angle is larger than others.
    
	\begin{figure}[t!]
		\centering
		\includegraphics[width=8.3cm]{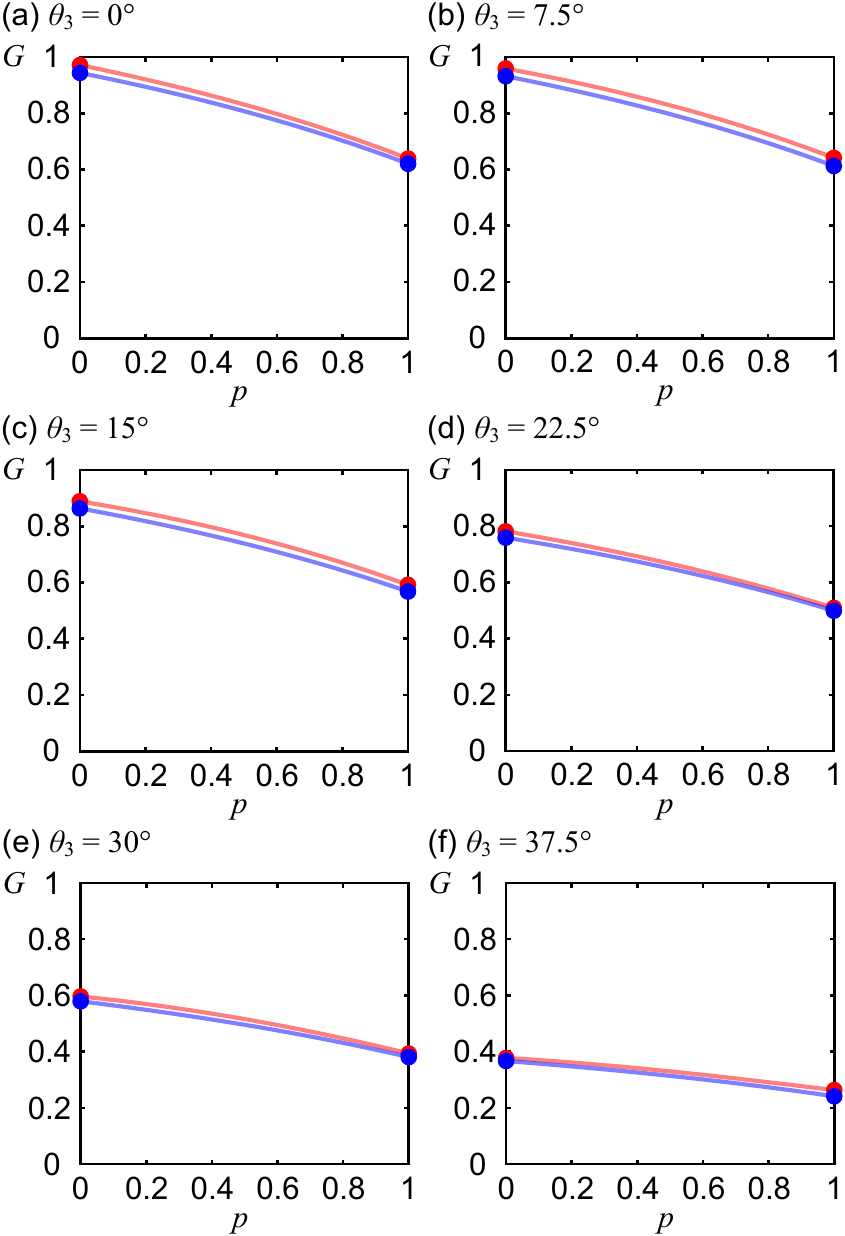}
		\caption{The $G$ values (light red curves) of the qutrit mixed state $\rho=(1-p)\left|\psi^+\right\rangle\left\langle\psi^+\right|+p\left|\psi'\right\rangle\left\langle\psi'\right|$ ($0\leq p\leq1$) after the phase damping operation with $\theta_3=$ 0°, 7.5°, 15°, 22.5°, 30° and 37.5° from the weighed averages of the count data, and the product values $G\left(\rho\right)G\left[\Phi\left(|\psi^+\rangle\langle\psi^+|\right)\right]$ (light blue curves). Measured $G$ and product values from the experiment are shown as red and blue dots respectively.
		}
		\label{qutritmix}
	\end{figure}
	
	The law of Eq. \ref{theorem} still holds for mixed initial states, which are statistical mixtures of different pure states. Weighed averages of count data can be used to simulate the mixed state scenario. We let the initial state be $(1-p)|\psi^+\rangle\langle\psi^+|+p|\psi'\rangle\langle\psi'|$ ($0\leq p\leq1$), where $|\psi'\rangle$ is the state when $\theta_2=$ 7.5°, and use the weighed averages of count data from $|\psi^+\rangle$ and $|\psi'\rangle$ as the new count data to simulate the mixed state scenario. Under six types of operations where $\theta_3=$ 0°, 7.5°, 15°, 22.5°, 30°, and 37.5°, corresponding to the six solid dots in Fig.~\ref{qutritd} (b), the calculated $G$ values (light red curves) and the product values (light blue curves) are shown in Fig.~\ref{qutritmix} when $p$ takes values from 0 to 1. The two curves are close to each other, verifying the factorization law with mixed input states.
	
	\section{Discussion and Conclusion}
	
	For qudits with a higher dimension, we can still use the path DOF, but the optical setup is more vulnerable to errors from the misalignment of optical devices, making the measured values deviate from theoretical ones. Nevertheless, we performed a four-dimensional experiment, projected the states after operation for ququad state tomography \cite{Thew2002}, and found the equation still agrees well for a certain operation and initial state: $G\left[\Phi\left(\left|\psi\right\rangle\left\langle\psi\right|\right)\right]\approx0.4349$, $G\left(\left|\psi\right\rangle\left\langle\psi\right|\right)G\left[\Phi\left(|\psi^+\rangle\langle\psi^+|\right)\right]\approx0.7406\times0.5849\approx0.4332$.
	
	There are other types of operations which satisfy the law. One example is that all the Kraus operators $K_n$ are multiplied by the same permutation matrix on the left, changing the order of the reference bases while keeping the coherence value. For example, a qutrit operation described by $\{K'_n\}$ is related to the Kraus operators $\{K_n\}$ of a GIO by
	\begin{align}
	    K'_n=\begin{pmatrix}0&K_{n,2}&0\\0&0&K_{n,3}\\K_{n,1}&0&0\end{pmatrix}=\begin{pmatrix}0&1&0\\0&0&1\\1&0&0\end{pmatrix}K_n.
	\end{align}
	But it does not hold when the permutation matrices for each $K_n$ are different. Also, some quantum operations cannot be described in the form above, but they are special to satisfy it as well. One example is the qubit amplitude decay channel
	\begin{align}
	    K_1=\begin{pmatrix}1&0\\0&\sqrt{1-\epsilon}\end{pmatrix},K_2=\begin{pmatrix}0&\sqrt{\epsilon}\\0&0\end{pmatrix},
	\end{align}
	where $0<\epsilon<1$, which describes the decay from the excited state to the ground state, whose correctness can be verified through calculation.
	
	In summary, we have presented and proved a factorization law for qudits under GIO, that using the \emph{G}-coherence measure, the coherence of the state after the operation can be factorized into the product of an initial state term and an operation term. To verify the law, we used an optical setup to test the qubit and qutrit case using a given set of initial states and operations. Our work provides an indirect method to measure the final coherence of quantum states after a specific kind of evolution, and would play an important role in the simplification of coherence measurement, as well as the discovery of other laws about quantum coherence. For example, there are other types of coherence metric, such as the convex-roof norm (See Ref. \cite{Qi2017} for more details) whose calculation method is uncertain yet. We can define $\Tilde{G}$ as the minimum statistical average $G$ of all the possible pure state combinations of $\rho$,
	\begin{align}
	    \Tilde{G}(\rho)=\min_{\{p_i,|\psi_i\rangle\}}\sum_i p_i G\left(\left|\psi\right\rangle\left\langle\psi\right|\right),
	\end{align}
	(For qubits $\Tilde{G}(\rho)=G(\rho)$ \cite{RMP,Qi2017}, but calculating this value for qudits is hard). There may be some factorization law for this coherence measure to be explored in the future, and this would deepen our understanding of quantum coherence. The possible applications of the coherence factorization law may be helping us design the effective coherence preservation schemes in quantum computation and quantum algorithm. For instance, many quantum algorithms have used $|+\rangle^{\otimes n}$ as input states with $|+\rangle=(|0\rangle+|1\rangle)/\sqrt{2}$,  which contain maximal coherence. However, if each qubit of the initial $n$-qubit state pass through a GIO channel such as bit and phase flip together, the output state do not contain maximal coherence. To preserve the coherence of output states, one can use the coherence factorization law and design proper schemes.
	
    \section*{Appendix: Qutrit State Tomography}
    \setcounter{equation}{0}
    \renewcommand{\theequation}{A\arabic{equation}}
    
    The Gell-Mann matrices are used in qutrit state tomography just as Pauli matrices in qubit case. For four-dimensional ququads, the matrices are the direct sums of two Pauli matrices. The original matrix $\Lambda_i$ ($i=1,2,\ldots,8$) can be found in Ref.~\cite{Gell-Mann,Thew2002}. They have 15 different normalized eigenvectors in total
	\begin{gather}
	    \left|\lambda_1\right\rangle=(1\quad0\quad0)^T,
	    \left|\lambda_2\right\rangle=(0\quad1\quad0)^T,
	    \left|\lambda_3\right\rangle=(0\quad0\quad1)^T,\nonumber\\
	    \left|\lambda_4\right\rangle=\frac{1}{\sqrt{2}}(-1\quad1\quad0)^T,
	    \left|\lambda_5\right\rangle=\frac{1}{\sqrt{2}}(1\quad1\quad0)^T,\nonumber\\
	    \left|\lambda_6\right\rangle=\frac{1}{\sqrt{2}}(\mathrm{i}\quad1\quad0)^T,
	    \left|\lambda_7\right\rangle=\frac{1}{\sqrt{2}}(-\mathrm{i}\quad1\quad0)^T,\nonumber\\
	    \left|\lambda_8\right\rangle=\frac{1}{\sqrt{2}}(-1\quad0\quad1)^T,
	    \left|\lambda_9\right\rangle=\frac{1}{\sqrt{2}}(1\quad0\quad1)^T,\nonumber\\
	    \left|\lambda_{10}\right\rangle=\frac{1}{\sqrt{2}}(-\mathrm{i}\quad0\quad1)^T,
	    \left|\lambda_{11}\right\rangle=\frac{1}{\sqrt{2}}(\mathrm{i}\quad0\quad1)^T,\nonumber\\
	    \left|\lambda_{12}\right\rangle=\frac{1}{\sqrt{2}}(0\quad-1\quad1)^T,
	    \left|\lambda_{13}\right\rangle=\frac{1}{\sqrt{2}}(0\quad1\quad1)^T,\nonumber\\
	    \left|\lambda_{14}\right\rangle=\frac{1}{\sqrt{2}}(0\quad\mathrm{i}\quad1)^T,
	    \left|\lambda_{15}\right\rangle=\frac{1}{\sqrt{2}}(0\quad-\mathrm{i}\quad1)^T.
	\end{gather}
	We define $\left\langle P_i\right\rangle=\operatorname{Tr}\left(\rho\left|\lambda_i\right\rangle\left\langle\lambda_i\right|\right)$ as the probability to be projected to state $\left|\lambda_i\right\rangle$. According to the eigenvalues and the corresponding eigenvectors, the average value of each Gell-Mann matrix is
	\begin{gather}
	    \left\langle\Lambda_1\right\rangle=\left\langle P_5\right\rangle-\left\langle P_4\right\rangle+0\left\langle P_3\right\rangle,
	    \left\langle\Lambda_2\right\rangle=\left\langle P_7\right\rangle-\left\langle P_6\right\rangle+0\left\langle P_3\right\rangle,\nonumber\\
	    \left\langle\Lambda_3\right\rangle=\left\langle P_1\right\rangle-\left\langle P_2\right\rangle+0\left\langle P_3\right\rangle,
	    \left\langle\Lambda_4\right\rangle=\left\langle P_9\right\rangle-\left\langle P_8\right\rangle+0\left\langle P_2\right\rangle,\nonumber\\
	    \left\langle\Lambda_5\right\rangle=\left\langle P_{11}\right\rangle-\left\langle P_{10}\right\rangle+0\left\langle P_2\right\rangle,
	    \left\langle\Lambda_6\right\rangle=\left\langle P_{13}\right\rangle-\left\langle P_{12}\right\rangle+0\left\langle P_1\right\rangle,\nonumber\\
	    \left\langle\Lambda_7\right\rangle=\left\langle P_{15}\right\rangle-\left\langle P_{14}\right\rangle+0\left\langle P_1\right\rangle,\nonumber\\
	    \left\langle\Lambda_8\right\rangle=\frac{1}{\sqrt{3}}\left\langle P_{1}\right\rangle+\frac{1}{\sqrt{3}}\left\langle P_{2}\right\rangle-\frac{2}{\sqrt{3}}\left\langle P_{3}\right\rangle,
	\end{gather}
	where the $+0\left\langle P_i\right\rangle$ terms mean $\left|\lambda_i\right\rangle$ is an eigenvector with zero eigenvalue, which is needed to calculate the projection probability from the count data. For example, $\left\langle P_5\right\rangle=C_5/\left(C_5+C_4+C_3\right)$, where $C_i$ is the photon count value when projecting to $\left|\lambda_i\right\rangle$. The density matrix can be reconstructed using Eq.~25 in Ref.~\cite{Thew2002}. For example,
	\begin{gather}
	    \rho_{12}=\left(\left\langle\Lambda_1\right\rangle-\mathrm{i}\left\langle\Lambda_2\right\rangle\right)/2,\nonumber\\
	    \rho_{13}=\left(\left\langle\Lambda_4\right\rangle-\mathrm{i}\left\langle\Lambda_5\right\rangle\right)/2,\nonumber\\
	    \rho_{23}=\left(\left\langle\Lambda_6\right\rangle-\mathrm{i}\left\langle\Lambda_7\right\rangle\right)/2.
	\end{gather}
	Calculating $\left\langle\Lambda_3\right\rangle$ and $\left\langle\Lambda_8\right\rangle$ is unnecessary if we only need its off-diagonal elements.

\begin{backmatter}
\bmsection{Funding} Innovation Program for Quantum Science and Technology (2021ZD0301400); National Natural Science Foundation of China (61725504, 11774335, 11821404, U19A2075, 11734015); Anhui Initiative in Quantum Information Technologies (AHY020100, AHY060300); Fundamental Research Funds for the Central Universities (WK2030380017); Open Funding Program from State Key Laboratory of Precision Spectroscopy (East China Normal University).

\bmsection{Disclosures} The authors declare no conflicts of interest.

\bmsection{Data availability} Data underlying the results presented in this paper are not publicly available at this time but may be obtained from the authors upon reasonable request.

\end{backmatter}

\end{document}